# Magnetization driven metal – insulator transition in strongly disordered Ge:Mn magnetic semiconductors.

O. Riss, A. Gerber, I.Ya. Korenblit,

Raymond and Beverly Sackler School of Physics and Astronomy
Tel Aviv University, Ramat Aviv, 69978 Tel Aviv, Israel,

A. Suslov

National High Magnetic Field Laboratory
1800 East Paul Dirac Drive, Tallahassee, Florida 32310-3706, USA,

M. Passacantando and L. Ottaviano

Dipartimento di Fisica, Università degli Studi dell'Aquila,
Via Vetoio, 67100 Coppito (AQ), Italy

We report on the temperature and field driven metal-insulator transition in disordered Ge:Mn magnetic semiconductors accompanied by magnetic ordering, magnetoresistance reaching thousands of percents and suppression of the extraordinary Hall effect by a magnetic field. Magnetoresistance isotherms are shown to obey a universal scaling law with a single scaling parameter depending on temperature and fabrication. We argue that the strong magnetic disorder leads to localization of charge carriers and is the origin of the unusual properties of Ge:Mn alloys.

PACS: 75.50 Pp, 75.47.-m, 73.50.-h

Interest in magnetic semiconductors was triggered by the development of spintronics in metallic magnetic materials and prospects to empower the semiconducting electronics by the spin-dependent degree of freedom. Compatibility with the existing silicon technology recently attracted much attention to the group-IV semiconductors doped with magnetic impurities. The novel materials appeared remarkably interesting and revealed a number of unusual properties not yet well understood. Huge positive magnetoresistance of hundreds to thousands of percent [1-4], reversal of magnetoresistance from positive to negative after annealing [4], and large Hall effect with non-monotonic field dependence [5-7] are among them. Interpretation of the data is difficult owing to the complicated structure of these alloys. As a result, almost no general relations were derived from the experiments, and only special models were proposed to explain the data in each case.

In this Communication we report on the high field magnetic and magnetotransport properties of several Ge:Mn samples produced by ion implantation. The material was found to pass from metallic-like to insulator-like state, either at low temperatures in zero field or under an applied field at elevated temperatures. We show that all the magnetoresistance data can be scaled on a universal curve with a single parameter $H_s$, which tends to zero at the paramagnet-to-ferromagnet transition. The same scaling procedure is applicable to the published data [4] obtained in samples produced by the molecular beam epitaxy, which indicates the generality of the scaling approach. We argue that establishment of an inhomogeneous magnetization landscape leads to localization of the charge carriers and, respectively, to a huge positive magnetoresistance and a total suppression of the extraordinary Hall effect.

Two Ge:Mn samples discussed here were fabricated by implanting $Mn^+$ ions into commercial single-crystalline Ge(100) wafers with resistivity of 40 – 57 Ωcm. $Mn^+$ ions were implanted with an energy of 100 keV at fluences of $1\times10^{16}$, and $2\times10^{16}$, that produce average volume concentration of Mn of about 2 and 4 at.% in the projected depth range of about 120 nm. During the implantation, the samples were held at 300 °C to avoid amorphization. Structure of the samples is strongly non-uniform and, depending on the concentration, contain diluted Mn, amorphous semiconducting Mn-rich nanoclusters and ferromagnetic metallic $Mn_5Ge_3$ clusters. Detailed structural, chemical and magnetic analyses of these samples can be found in Refs [8,9]. The 2% sample has a Hall bar geometry, while the 4% sample was measured in the Van der Pauw configuration.

Resistivity of the 2% sample measured as a function of temperature in zero and 16 T field is plotted in Fig.1. The zero field resistivity is metallic ($d\rho/dT > 0$) above 30 K, it can be well fitted by the power-law temperature dependence: $\rho = \rho_0 + \alpha T^{3/2}$ up to 180 K. Such behavior is typical for the phonon scattering of non-degenerate carriers with a temperature independent concentration. Similar behavior was reported in Ref. [1]. A step-like metal – insulator transition takes place below 30 K with resistivity increasing by two orders of magnitude, followed by a more moderate increase below 8 K. In contrast with the zero field case, the resistivity measured under high magnetic field (16 T data in Fig.1) shows an insulator-like temperature dependence ($d\rho/dT < 0$) in the entire range up to room temperature. The result is the same both for the field cooled (FC) and the zero field cooled (ZFC) protocols.

Magnetization of the sample was measured in a high field SQUID at several temperatures. The magnetization values reached at 7 T are shown by stars in Fig.1. The magnetization increases sharply below 30 K similar to the found in other Ge:Mn samples [9]. Maximum of susceptibility was found in precipitate-free Ge:Mn [10] in the same temperature range, indicating the onset of the ferromagnetic order. However, we did not detect the remanent magnetization in zero field at low temperatures. This suggests that either the ferromagnetic state is highly non-uniform, or that the ferromagnetic order is established only in large but finite domains with zero macroscopic magnetization. Remarkably, the rise of magnetization coincides with the steep enhancement of resistivity and marks the transition between the magnetically disordered metallic state at elevated temperatures and the magnetically ordered and electrically insulating state at low temperatures.

The field dependence of the resistivity was measured at different temperatures up to 16.5 T at TAU, and up to 33.3 T at NHMFL at Tallahassee. Results for the 2% sample are typical and presented in Fig. 2. Huge positive magnetoresistance is observed down to the insulator range temperatures with the normalized ($\rho(33T)/\rho(0T)$) values of 220% at room temperature and 3200% at 40K [11]. An intriguing feature of the phenomenon is the existence of a cross-over field, separating the metal-like phase ($d\rho/dT > 0$) at low fields from the insulator-like ($d\rho/dT < 0$) phase at high fields. The cross-over field ( see Fig. 2) is roughly independent of temperature above 100 K, and appears in all samples studied. To the best of our knowledge such field driven metal – insulator transition has not been observed in Ge doped by non-magnetic impurities.

Hall resistivity measured in 2% sample at a variety of temperatures is shown in Fig.3. Above 200 K the signals contain both the ordinary (OHE) and the extraordinary Hall effect (EHE) contributions, and remind the usual form: $\rho_H = R_0 B + \mu_0 R_{EHE} M$, where $R_0$ and $R_{EHE}$ are the ordinary and extraordinary Hall effect coefficients respectively. The field dependence of the EHE term at these temperatures can be fitted by the Langevin function with the saturated value of magnetization of about 100 $\mu_B$. The effect can be due to asymmetric scattering of conducting electrons by magnetic clusters, similar to that observed in heterogeneous systems with ferromagnetic clusters embedded in non-magnetic metallic matrices [12]. The unusual behavior of the Hall signal is emphasized at low temperatures, e.g. at 8 K (see inset Fig. 3). The Hall signal increases sharply at low fields, reaches a peak, decreases and gradually approaches a positive linear slope corresponding to the ordinary Hall effect component. The EHE is entirely suppressed at high magnetic field. Similar behavior was probably observed in other dilute magnetic semiconductors Ge:Cr [6] and (In,Mn)Sb [7]. Non-monotonic field dependence of the extraordinary Hall coefficient in metallic magnetic materials has been observed in granular and multilayer systems and was proposed [13] to be linked to the giant magnetoresistance (GMR) effect. However, to the best of our knowledge, the total suppression of EHE by magnetic field has not been identified so far. At 4.2 K only linear in the field Hall signal is found, which implies that the EHE is zero in the insulator phase both below the transition temperature at low fields or at high temperatures beyond the cross-over field.

As shown in Fig. 2 the magnetoresistance tends to saturate in strong fields, the saturation field being larger at higher temperatures. This hints at a scaling dependence of magnetoresistance on *H* and *T*. We succeeded to rescale the magnetoresistance $\rho(H) - \rho(0)$ by a factor $A(T,x)$ and the field by a factor $H_s(T,x)$, where $x$ is Mn concentration, in such a way that all the experimental points at $T \geq 15$ K collapse to one curve (Fig.4). In the limit of strong fields $H \gg H_s$ this curve extrapolates to unity, which implies that $A(T,x)$ equals the full change of the resistivity, $\rho(\infty) - \rho(0)$, where $\rho(\infty)$ is resistivity in the limit of an infinite field. We, thus, obtain

$$\frac{\rho(H) - \rho(0)}{\rho(\infty) - \rho(0)} = F\left(\frac{H}{H_s}\right) \tag{1}$$

The data collapse is accurate above 40 K up to room temperature, but at lower temperatures, when the behavior of the zero-field resistivity starts to change from metallic to insulating, the scaling is good only for fields H ≤ 20 T at T = 20 K and H ≤ 10 T at T = 15 K.

The temperature dependence of the parameter $\rho(\infty)$, shown in Fig.1, is insulator-like in the entire temperature range and is consistent with the experimental data obtained at 16 T. $H_s(T)$ for 2% and 4% samples as shown in Fig. 5. $H_s(T)$ decreases to zero at some temperature $T^*$ close to 15 K. At temperatures above $T^*$ the field $H_s$ varies linearly with temperature as: $H_s = a + bT$, where $a$ = 11 T and $b$ = 0.056 T/K for both samples. At sufficiently high temperatures $(T > a/b)$ $H_s \propto T$, and the magnetoresistance is a function of $H/T$, similar to magnetization of non-interacting localized spins or clusters. One concludes, therefore, that $H_s$ at $T > T^*$ behaves as the inverse susceptibility, and the magnetoresistance appears to be proportional to the magnetization, at least at H < $H_s$. The scaling function $F(H/H_s)$ can be fitted accurately by a simple function

$$F(h) = 1 - \exp(-h) \tag{2}$$

For fields $H \leq H_s$ the scaling function can also be approximated by the Langevin function $L(h)$ with $h = 3H/H_s$. At high temperatures, where $H_s \propto T$, the last approximation gives for the magnetic moment $m$ of the clusters: $m = 3k_B/b \approx 80\mu_B$, where $k_B$ is the Boltzmann constant. This value is close to the moment of clusters estimated above from the EHE data at high temperatures.

In addition to our data we tested the applicability of the scaling functions (1) and (2) to the results published in Ref [4]. The samples reported in this paper were, unlike ours, fabricated by the molecular beam epitaxy, have a very different morphological structure and include strings of ferromagnetic nanocolumns. Nevertheless the fit, as seen from Fig.4, is very good. $H_s$ grows linearly at high temperatures (see Fig.5) similar to our data. Thus, the scaling law (1) and (2) appears to be a general feature of the magnetoresistance in the metallic-like range of Ge:Mn, while the scaling field $H_s$ depends on the preparation details and temperature.

Several attempts were made to explain a huge magnetoresistance in Ge:Mn. Orbital effects enhanced by the presence of conducting nanocolumns were evoked in Ref. [1]. In our samples the Hall angle is small even in strong fields, which makes this mechanism ineffective. Positive magnetoresistance of the order of 100% in $Ge_{1-x}Mn_x$ epitaxial films was explained in Ref. [4] by scattering off the spatial fluctuations of the atomic spin concentrations (see also Ref.[14]). This mechanism cannot explain the magnetoresistance of thousands of percent as well as the M-I transition observed by us. However, the change of the carrier mobility [4, 14], which does not follow the scaling law, can lead to the small deviations from scaling seen in Fig. 4.

Both the large positive magnetoresistance at high temperatures and the strong increase of the resistivity with decreasing temperature in the quasi-ferromagnetic phase, show that the resistivity increases with the increase of the magnetization. As noted above, the hole concentration is independent of temperature in the metallic range, which implies that the impurity levels (acceptors) are very shallow, and a uniform exchange shift of the bottom of the valence band cannot change the hole concentration. However, because of the non-uniform distribution of Mn dopants, the magnetization, spontaneous or induced by the magnetic field, strongly fluctuates on a scale much larger than the interatomic distance, and so does the exchange shift of the top of the valence band. The fluctuating part of the exchange shift $\Delta(r)$ is:

$$\Delta(r) = Jm(r)\sigma = Jx(r)\sigma <S_z> \quad (3)$$

where $J$ is the exchange coupling energy, $m(r)$ is the fluctuating part of the local magnetic moment, $x(r)$ is the fluctuating in space concentration of Mn, $\sigma = \pm 1/2$, and $<S_z>$ is the thermodynamically averaged projection of the local spin. This forms a complicated landscape of the valence band with hills and valleys serving as traps for holes. With the rise of magnetization the traps get deeper, the number of conducting holes decreases, and the resistivity increases. At sufficiently high fields or low temperatures, the Fermi level of carriers descends below the percolation threshold and the conductance changes its character from metal-like to insulating. One can then expect that the magnetoresistance should be smaller in samples with higher zero-field resistivity which are closer to the metal-insulator (M-I) transition. This prediction seems to be correct when we compare our data and that in Ref. [1] to magnetoresistance observed in the samples of Ref [4]. These samples are on the verge of the M-I transition and their magnetoresistance is an order of magnitude smaller than reported here and in Ref. [1]. On the other hand, in samples with weak disorder the magnetization fluctuations cannot form traps sufficient to localize the holes, and the MR should be small. This is in agreement with the observation of Ref [4] that in annealed samples the magnetoresistance is small and even negative.

Inset of Fig.4 presents the EHE component of the Hall signals measured in 2% and 4% samples as a function of $H/H_s$ for different temperatures. At all temperatures the EHE contribution reaches a maximum at fields $H \approx H_s/2$ and disappears at $H >> H_s$ when the system is in the high field insulating state. $H_s$ appears to serve as a characteristic field for the extraordinary Hall effect as well, which links the magnetoresistance and Hall effect behavior.

Suppression of the EHE in the insulating state might be consistent with a model of hopping conduction in systems with strong Hund's rule [15], which predicts that EHE tends to zero when the magnetization saturates. At this moment such interpretation is yet premature.

To summarize, Mn doped Ge enters the insulating state below 20-30 K at zero field and under high applied magnetic field at higher temperatures. The transition to the low-temperature insulating state is accompanied by the transition into a highly disordered ferromagnetic phase. Field driven transition into the insulating state is the origin of a huge positive magnetoresistance of thousands of percent. The extraordinary Hall effect is suppressed in both the low-temperature and the high-field insulating states. All magnetoresistance isotherms in the metallic-like temperature range can be scaled by a universal function with a single scaling field parameter that depends on temperature and fabrication. The scaling is general and is applicable to samples prepared by different methods and having different disordered microscopic structures. The same scaling field parameter characterizes also the magnetic field dependence of EHE. We argue that the transition to the insulating state is related to a strongly inhomogeneous distribution of magnetic moments that leads to localization of conducting holes. Thus, it is the strong magnetic disorder which is responsible for the colossal magnetoresistance and other unusual properties of Ge:Mn alloys. Universality of the scaling is remarkable and should serve the test tool for microscopic theories of magnetotransport in magnetic semiconductors.

We thank Giuliana Impellizzeri and Francesco Priolo from MATIS-INFM-CNR for the Mn ion implantation of the Ge samples studied here, and Eun Sang Choi for help with magnetization measurements. This work was supported by the Israel Science Foundation grant No. 633/06. NHMFL is supported by NSF Cooperative Agreement No.DMR-0654118, the State of Florida and the DOE.

**Figure Captions.**

Fig.1. Temperature dependence of: (i) zero field resistivity (○); (ii) resistivity under 16 T field (●); (iii) extrapolated resistivity at an infinite field (×) (see the definition in text); (iiii) magnetization at 7T applied magnetic field (stars) for Ge:Mn sample with 2% Mn doping.

Fig.2. Resistivity of the 2% sample as a function of applied magnetic field at different temperatures. The cross-over region between the metallic-like and insulating phases is indicated by the circle.

Fig.3. Hall resistivity measured in the 2% Mn sample at different temperatures. Inset: zoom of the 4.2 K, and 8 K data at low fields.

Fig.4. Scaling of the 2% and 4% samples magnetoresistance measured at different temperatures. Solid line is the fit calculated by Eq.2. Data marked by large symbols (+, ○, ×, ∗) were taken from Ref.4. Inset: the EHE resistivity component of 2% and 4% samples as a function of $H/H_s$. Location of the maximum is roughly independent of temperature.

Fig.5. Scaling parameter $H_s$ for 2% (circles) and 4% samples (crosses) as a function of temperature. Dotted line is guide for the eyes. Drop of $H_s$ towards zero indicates the phase transition to the disordered ferromagnetic state. Triangles indicate the values of $H_s$ calculated for data taken from Ref. 4.

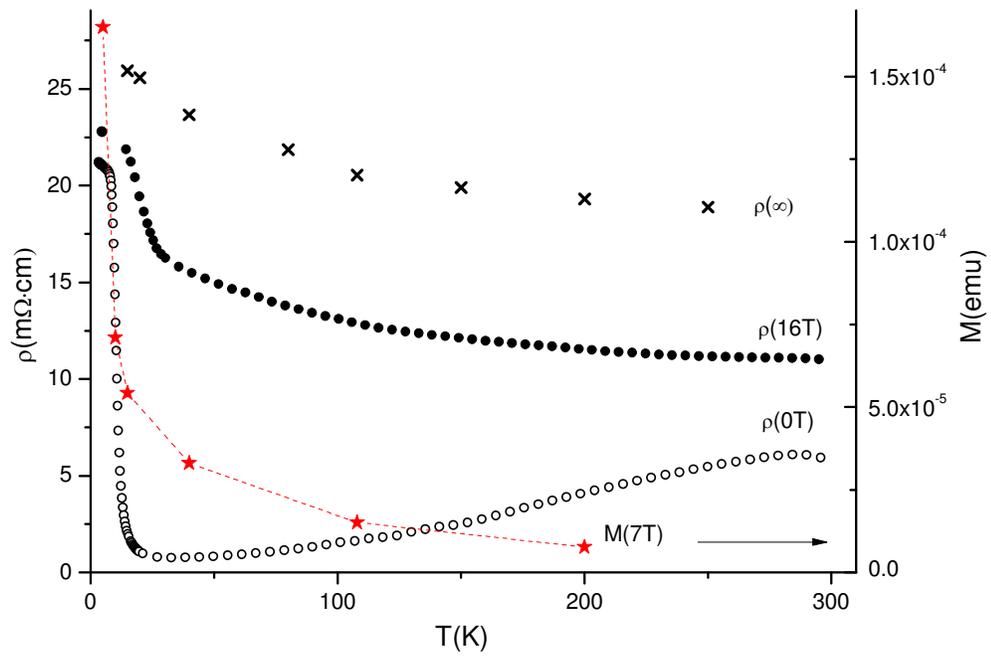

Fig. 1

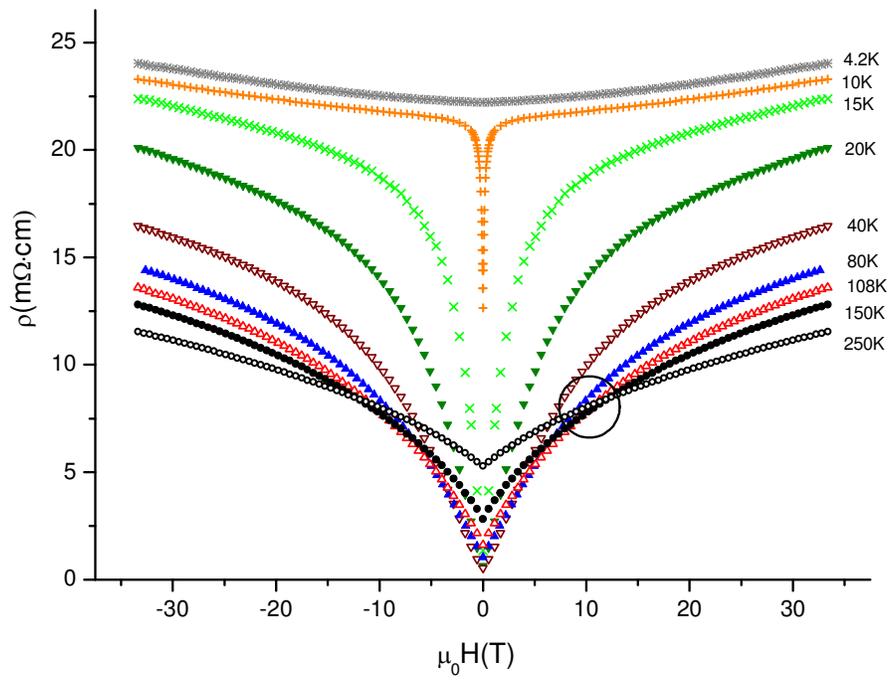

Fig.2

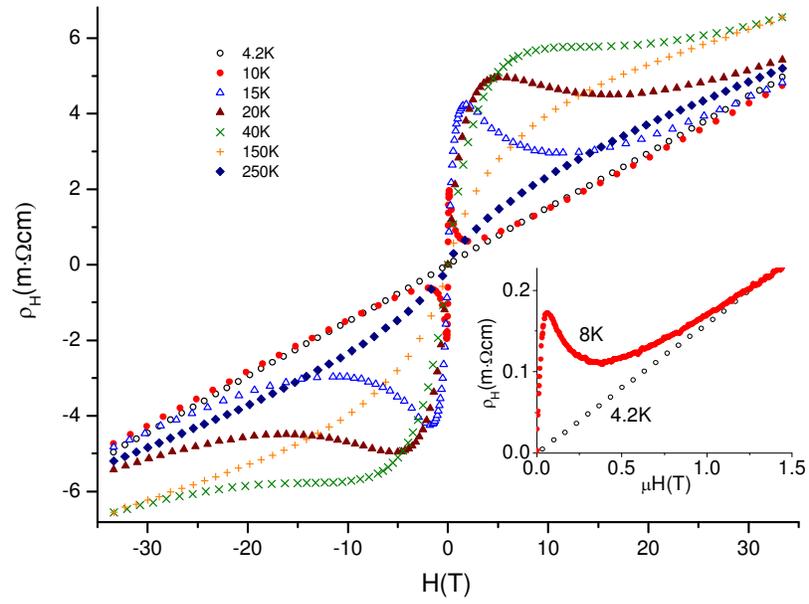

Fig. 3

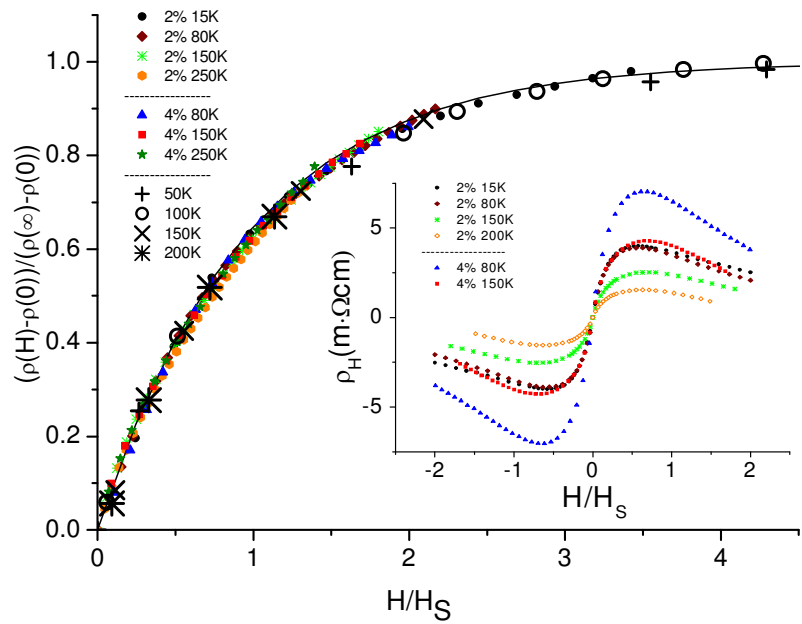

Fig. 4

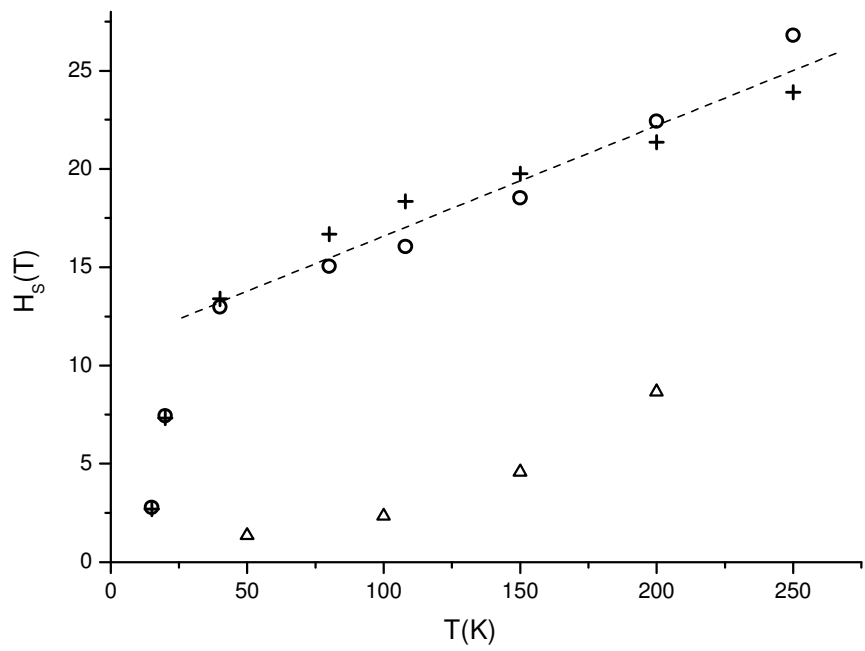

Fig. 5